\documentclass[a4paper,11pt]{article}
\usepackage{pos}

\usepackage{natbib}
\title{Analysis of EAS-like events detected by the Mini-EUSO telescope}
 \ShortTitle{EAS-like events in Mini-EUSO data}

\author*[a,b,c]{Matteo Battisti}
\author[b,c]{Mario Bertaina}
\author[b,c]{Enrico Arnone}
\author[c]{Gaetano Sammartino}
\author[c]{Giulia Pretto}

\affiliation[a]{Universit\'e de Paris, CNRS, AstroParticule et Cosmologie, F-75013 Paris, France,}
\affiliation[b]{Istituto Nazionale di Fisica Nucleare - Sezione di Torino, Italy, }
\affiliation[c]{Dipartimento di Fisica, Universita' di Torino, Italy}

\onbehalf{for the JEM-EUSO Collaboration\\[-1mm]{\normalsize \normalfont (a complete list of authors can be found at the end of the proceedings)}}
\emailAdd{battisti@apc.in2p3.fr}
\emailAdd{mbattist@to.infn.it}
\emailAdd{matteo.battisti@edu.unito.it}

\abstract{The Mini-EUSO telescope is the first space-based detector of the JEM-EUSO program. It was launched for the International Space Station on August 22$^{nd}$, 2019 to observe from the ISS orbit ($\sim$420~km altitude) various phenomena occurring in the Earth’s atmosphere through a UV-transparent window located in the Russian Zvezda Module. The dimension of the window defines and constrains the dimension of the optics, based on a set of two Fresnel lenses of 25~cm diameter each, almost two orders of magnitude smaller than the system foreseen for a larger space-based detector, like the original JEM-EUSO detector or the future POEMMA. As a consequence, the energy threshold of Mini-EUSO is very high, above $10^{21}~$eV. Nevertheless, a series of events that resemble the shape and the time duration of EAS-induced events have been detected in Mini-EUSO data. This contribution presents the most interesting cases, showing that the vast majority of the EAS-like events can be traced back to ground sources repeatedly flashing and triggered many times by Mini-EUSO. Some non-repeated EAS-like events are also present. In these cases, it is possible to exclude their cosmic origin through the comparison with simulated events. 
Since it is clear that those events can not be originated by a UHECR, we decided to rename them "Short Light Transients" or SLTs.
Finally, it was possible to associate some of the SLTs with atmospheric activity. This analysis confirms the validity of the JEM-EUSO detection principle and shows that it is possible for a space-based detector to distinguish between events induced by UHECRs and events with a different origin.}

\ConferenceLogo{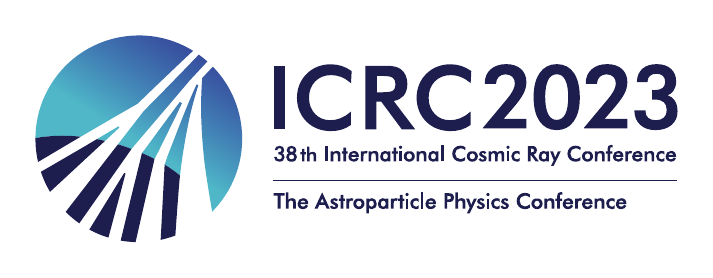}

\FullConference{%
38th International Cosmic Ray Conference (ICRC2023)\\
  26 July - 3 August, 2023\\
  Nagoya, Japan}

\begin{document}
\maketitle

\section{Mini-EUSO mission}
\begin{figure}[hbtp]
\centering
\includegraphics[width=.98\textwidth]{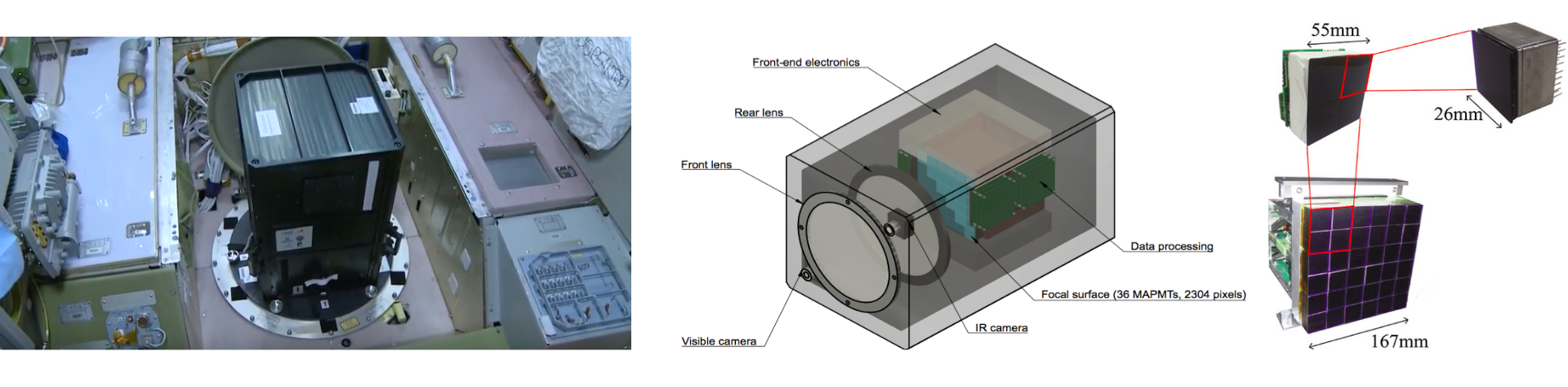} 
\caption{\textbf{Left:} Mini-EUSO installed on the UV-transparent window during a data-taking session. \textbf{Middle:} Schematic view of the instrument. \textbf{Right:} The focal plane, made of an array of MAPMTs, for a total of 2304 pixels arranged in a $48\times48$ matrix.}
\label{fig:Mini-EUSO_FlightModel}
\end{figure}

Mini-EUSO~\cite{minieuso} (Fig.~\ref{fig:Mini-EUSO_FlightModel}) can be thought of as a compact version of the original JEM-EUSO telescope~\cite{JEM-EUSO_telescope}. It is the first detector of the JEM-EUSO program to observe the Earth from the ISS and to prove the observational principle of the original JEM-EUSO detector from space.  The optical system is made of a set of two Fresnel lenses of 25~cm diameter each, the focal plane is made of an array of 6$\times$6 MAPMTs, for a total of 2304 pixels arranged in a 48$\times$48 matrix.

%For this reason Mini-EUSO has been designed to detect a photon count rate per pixel per unit time from diffuse sources (nightglow, clouds, cities, etc..) similar to that expected from a large mission in space such as K-EUSO or POEMMA, or the original JEM-EUSO mission.

Mini-EUSO has been designed to detect a photon count rate per pixel per unit time from diffuse sources (nightglow, clouds, cities, etc..) similar to that expected from a large mission in space such as POEMMA~\cite{poemma} or the original JEM-EUSO mission. The pixels footprint is $\sim$100 times larger than the original JEM-EUSO detector,  resulting in $\sim$6.3~km on ground, to compensate for the smaller optical system constrained by the dimension of the UV transparent window. The consequence of the relatively small optical system and the large field of view of each pixel is that the energy threshold for the detection of UHECRs is above $10^{21}$~eV~\cite{Mario_exposure_EPJ}. 

The system works in single photon-counting mode. Photon counts are summed every 2.5 $\mu$s, forming what is called a D1 GTU (Gate Time Unit). %The Mini-EUSO data acquisition system takes data with two other time resolutions of 320 and 40.96 ms called D2 and D3 respectively. These other time resolutions will not be considered in this analysis.
Once a trigger is issued, 128 D1 GTUs are stored. We refer to the collection of 128 GTUs as a "packet" or as an "event". 
%The time resolution is 2.5~$\mu$s, identical to the one foreseen for the original JEM-EUSO detector. 
A detailed description of the Mini-EUSO trigger logic is reported in~\cite{belov-trigger}. %, while the on-board performance of the trigger system is summarized in~\cite{matteo-trigger}.
In brief, the algorithm searches for a signal integrated over 8 consecutive GTUs above 16
standard deviations from the average in any pixel of the focal surface. The average and standard
deviation are updated every 320~$\mu$s to avoid triggering on slowly changing background. In case of a trigger, 128 D1 GTUs (a packet) are stored in memory, 64 GTUs before the trigger and 64 GTUs after it.

The system has an internal safety mechanism that prevents damage to the MAPMTs by reducing their collection efficiency and gain in the presence of very bright light. The status of reduced efficiency is called \textit{cathode 2}. The nominal working condition is instead called \textit{cathode 3}. 

The analysis presented here is performed on the data gathered between October 2019 and August 2021, for a total of $\sim$120 hours of observation and over 120,000 D1 triggered events. %After each session a small portion of the data gathered is immediately downlinked to ground ($\sim$40 minutes of data, that correspond to $\sim$20\% of the acquired data, usually the first and last part of the session) while the remaining part is stored on 512 GB USB Solid State Disks (SSDs) and delivered by the astronauts upon their return to Earth.
%Pouches with 25 SSDs are returned to Earth every 6-12 months.

\section{The analysis}
\begin{figure}[hbtp]
\centering
\includegraphics[width=.98\textwidth]{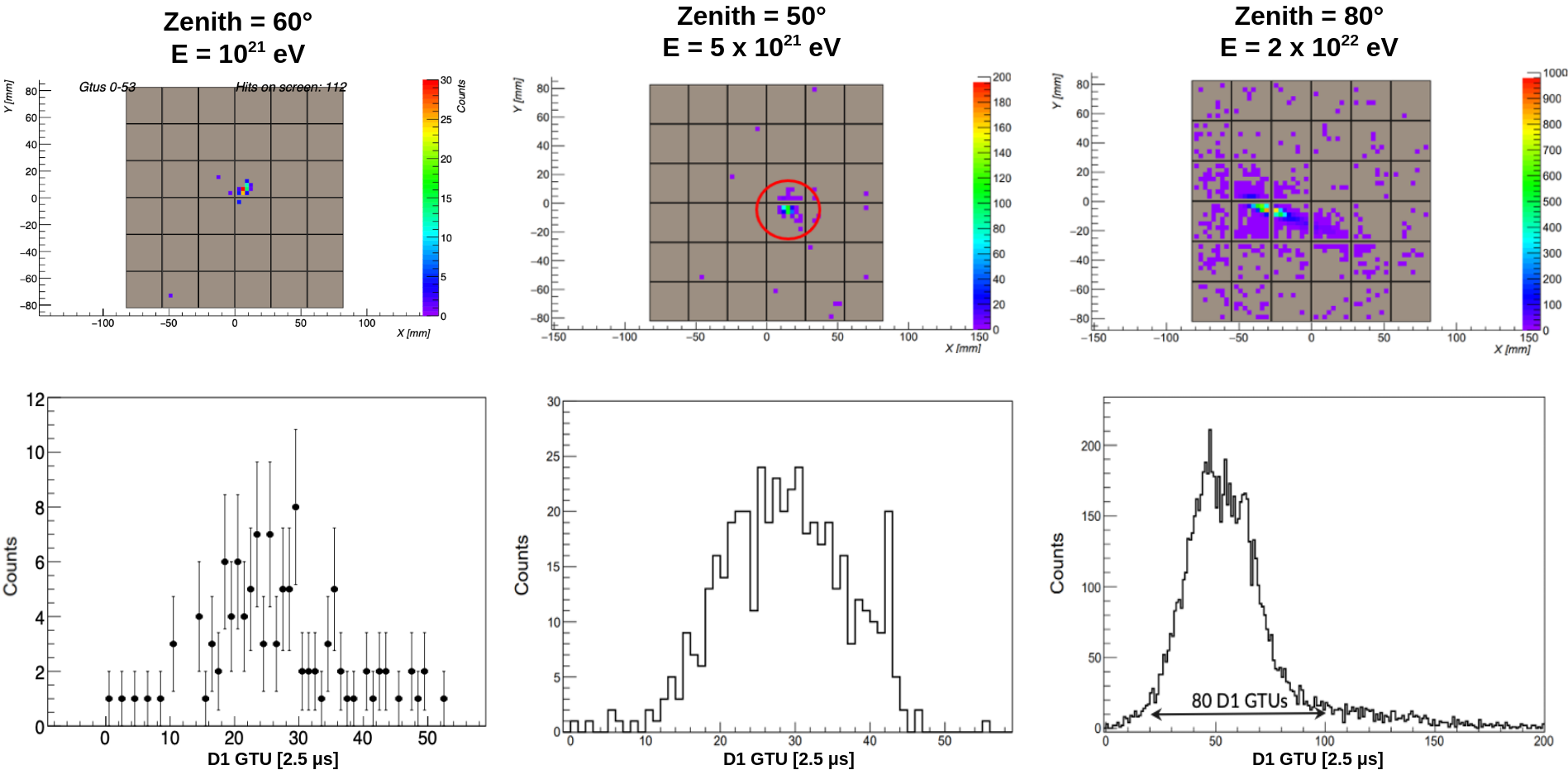} 
\caption{ESAF~\cite{ESAF} simulation of protons with different energies and zenithal angles. Given its high energy threshold, a shower of 10$^{21}$~eV \textbf{(Left)} is at the edge of the Mini-EUSO triggering capabilities. Around $5\times10^{21}$~eV the signal of a 50$^\circ$ shower (\textbf{Middle}) is clearly visible and lasts for $\sim$30~GTUs ($\sim$75~$\mu$s). The signal is truncated when the shower reaches the ground. At 80$^\circ$ zenith angle (\textbf{Right}, energy $2\times10^{22}$~eV) the lightcurve is not truncated and the signal lasts for $\sim$80~GTUs ($\sim$200~$\mu$s). }
\label{fig:ESAF_simulation_Mini_EUSO}
\end{figure}

An extensive air shower (EAS) would appear in Mini-EUSO as a signal persisting in a pixel for at least $\sim$8 GTUs and eventually moving to neighboring pixels, with a light profile that matches a bi-gaussian shape, with a faster rising and a slower decay that can be abruptly truncated for vertical events when the shower reaches the ground. Three examples of EAS simulations are shown in Fig.~\ref{fig:ESAF_simulation_Mini_EUSO} for different energies and zenithal angles. The main goal of the analysis presented here is the research and identification over the entire dataset of signals with similar characteristics.

The analysis has been performed through the following steps.

\begin{figure}[h]
\centering
\includegraphics[width=.99\textwidth]{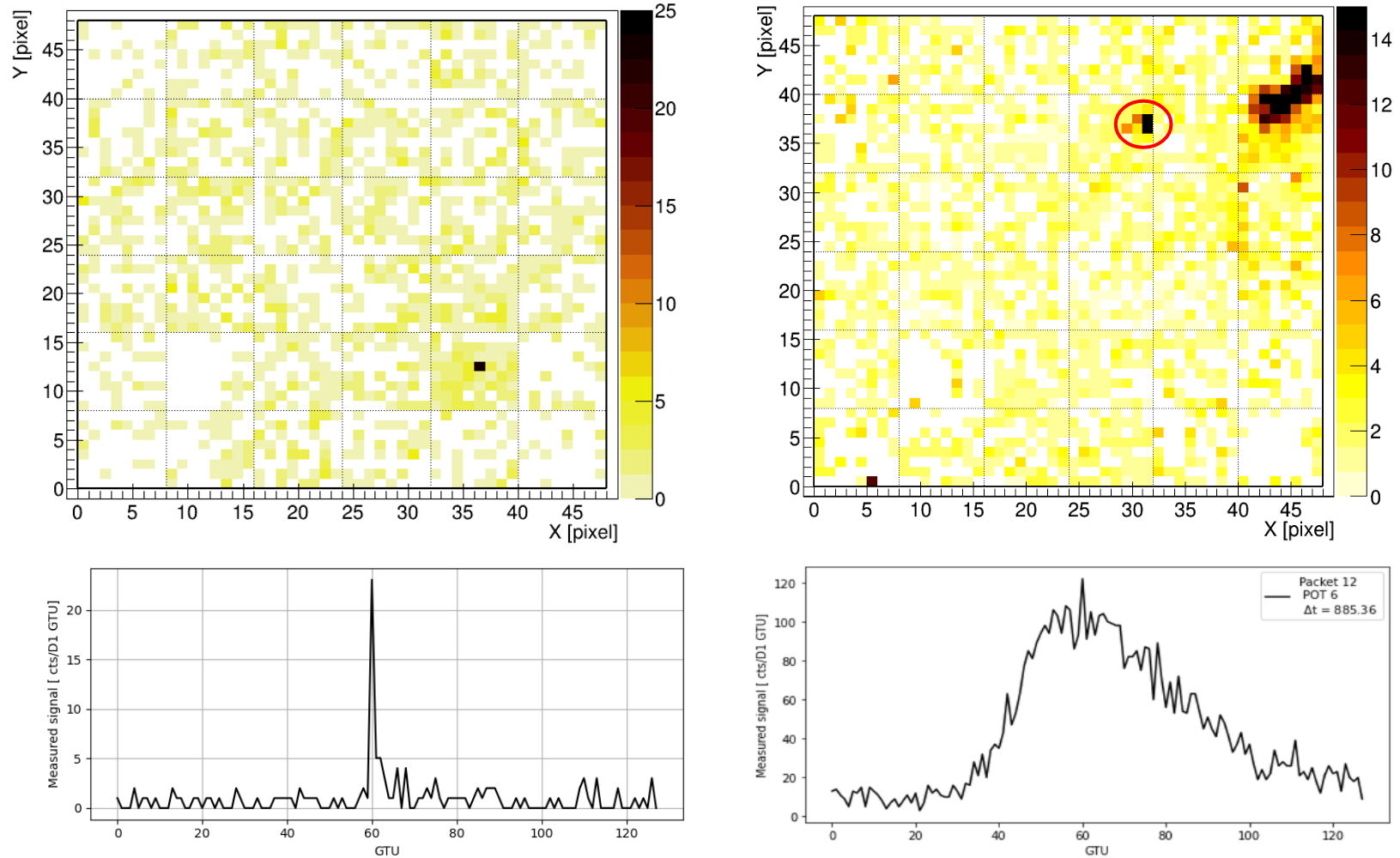}
\caption{\textbf{Left:} Example of a direct cosmic ray. The trigger is caused by a low energy cosmic ray impinging on the detector and leaving a bright signal in one pixel. The bottom plot shows the lightcurve of the brightest pixel. \textbf{Right:} Example of a EAS-like event (or, more properly, a "Short Light Transients", SLT). The event has been detected off the coast of Sri Lanka (the bright area on the right of the focal plane). It appears in a small cluster of pixels and shows a bi-gaussian lightcurve, with a faster rise and a slower decay. The lightcurves shown in this work are the sum of all the pixels over threshold (POT in the legend) in the packet (6 in this case). }
\label{Fig:DCR_and_EAS_like}
\end{figure}

\begin{itemize}
    \item \textbf{Apply the trigger logic.} The entire dataset has been analyzed through an offline algorithm that mimics the online trigger logic implemented in the detector. The details of the algorithm are presented in ~\cite{matteo-trigger}. The aim is to identify for each event all the pixels over threshold, since in the following steps of the analysis only their lightcurve and their position on the focal plane are taken into account. The packets without a recognized pixel over threshold are discarded. They account for $\sim$10\% of the entire database and are due to the overestimation of the threshold values by the offline algorithm\footnote{The values of the background required to compute the thresholds for each pixel have to be estimated from the data since Mini-EUSO does not store them. The amount of data represented by the thresholds would be, in fact, an additional 50~GB per hour; such an amount of data would be impossible to handle for a space-based experiment.}. %As mentioned above, the thresholds are updated every 320~$\mu$s and it is therefore not possible to store this huge amount of data. The offline algorithm estimates their values from the data.}. 
    Indeed, a bright signal is present in most of these events and it is recognized if the threshold is artificially lowered at 12~$\sigma$ (see~\cite{matteo-trigger} for details).%Only the pixels over the threshold are considered in the next steps of the analysis.
    \item \textbf{Discard the direct cosmic ray events.} Triggers caused by low energy cosmic rays directly impinging on the detector are excluded from the analysis. These events are usually characterized by a signal that reaches the maximum in 1 or 2 D1 GTUs and then shows an exponential decay, as shown in Fig.~\ref{Fig:DCR_and_EAS_like}, left. Three simple cuts on the shape, duration, and number of peaks are enough to identify and discard the direct cosmic ray hits. Also, the triggers issued by 12 noisy pixels (0.5\% of the total number of pixels) are excluded from the analysis. Direct cosmic rays and electronic-induced trigger account for $\sim$60\% of the entire database.
    \item \textbf{Visual Analysis.} The remaining $\sim$30\% of the database (almost 37,000 events) is manually checked. This subset mainly contains triggers issued by atmospheric events, by the recovery to \textit{cathode 3} after the \textit{cathode 2} switch, and by the light of the rising sun. The analysis is focused on the research of two types of events: short, not-repeated flashes and flasher signals. Events with a duration $\leq200~\mu$s were considered candidates of EAS-like events. As it will be shown if Sec.~\ref{Sec:SLT} we can exclude that those events are originated by EASs in the atmosphere and we prefer to call them "Short Light Transients" or SLTs.  An example is shown in Fig.~\ref{Fig:DCR_and_EAS_like}, right. Events triggered many times as the ISS passes by are considered to be produced by ground flashers. These events sometimes have lightcurves that recall at first sight the ones shown in Fig.~\ref{fig:ESAF_simulation_Mini_EUSO}.
\end{itemize}

\subsection{Ground flashers}
\label{Flashers}
\begin{figure}[h]
\centering
\includegraphics[width=.99\textwidth]{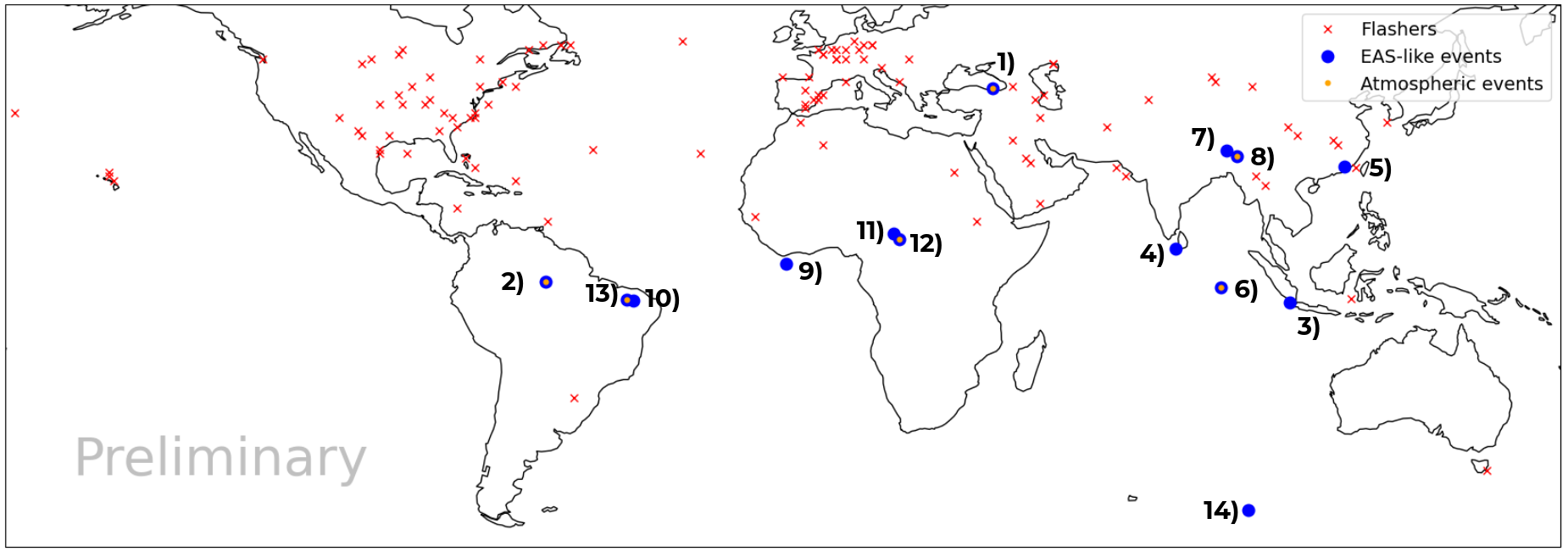}
\caption{\textbf{Red:} Approximate location of 108 (out of 561) ground flashers detected by Mini-EUSO. The vast majority of the flashers are located close to the coastlines or near highly populated areas. The origin of the 4 flashers found over the ocean is still under study. \textbf{Blue:} Position of the 14 SLT events. The numbers refer to the lightcurve shown in Fig.~\ref{Fig:EAS_like_events__lightcurves}. \textbf{Magenta:} Atmospheric events detected within a few ms after an SLT. See the text for more details.}
\label{Fig:Map}
\end{figure}

\begin{figure}[h!]
\centering
\includegraphics[width=.99\textwidth]{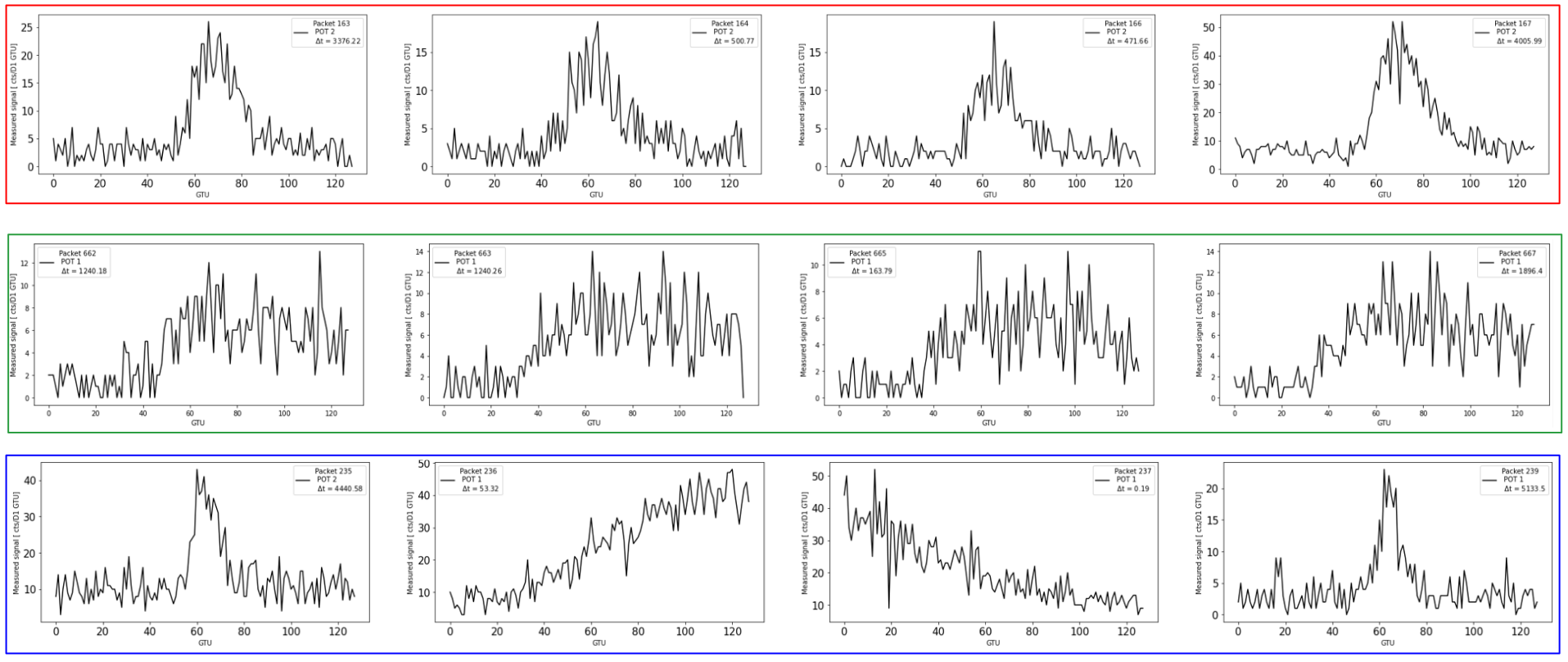}
\caption{Typical flashers' lightcurve. Each plot represents the sum of the lightcurves of the pixels over threshold (POT in the legend) in the event. The periodic signal is triggered many times, the differences between different packets are due to the statistical fluctuations, the different efficiency of the pixels and the number of pixels over threshold.  \textbf{Top, Red:} The signal lasts for a few tens of GTUs. The resulting lightcurve can be similar to the EASs simulations shown in Fig.~\ref{fig:ESAF_simulation_Mini_EUSO}. \textbf{Middle, Green:} The signal is relatively weak and longer than 320~$\mu$s, therefore it is not fully contained in a packet. \textbf{Bottom, Blue:} The same flasher alternates between short pulses (first and last plot) and long pulses lasting $\sim$0.5~ms that are triggered in two consecutive packets (second and third plot).}
\label{Fig:Flasher_lightcurve}
\end{figure}

Thanks to the visual inspection of the events, we were able to identify 561 different ground flashers. The location of 108 of them (coming from 8 different sessions) is reported in Fig.~\ref{Fig:Map} in red. Every one of these flashers has been triggered several times (at least 3, but usually up to 15 or 20 times) in the $\sim$50~s that a point on the ground stays inside the Mini-EUSO field of view.  The signal is usually confined in a small cluster of pixels, while the intensity and duration can change a lot from case to case, as exemplified by Fig.~\ref{Fig:Flasher_lightcurve} and can, sometimes, resemble the signal expected by a UHECR. Their anthropogenic origin is however confirmed whenever the same event is triggered several times and they are never misidentify for EAS-induce signals, despite their large number.  Most likely the flasher signals are produced by blinking lights on ground, usually located near airports, ports or cities.

\subsection{Short Light Transients}
\label{Sec:SLT}
\begin{figure}[h!]
\centering
\includegraphics[width=.99\textwidth]{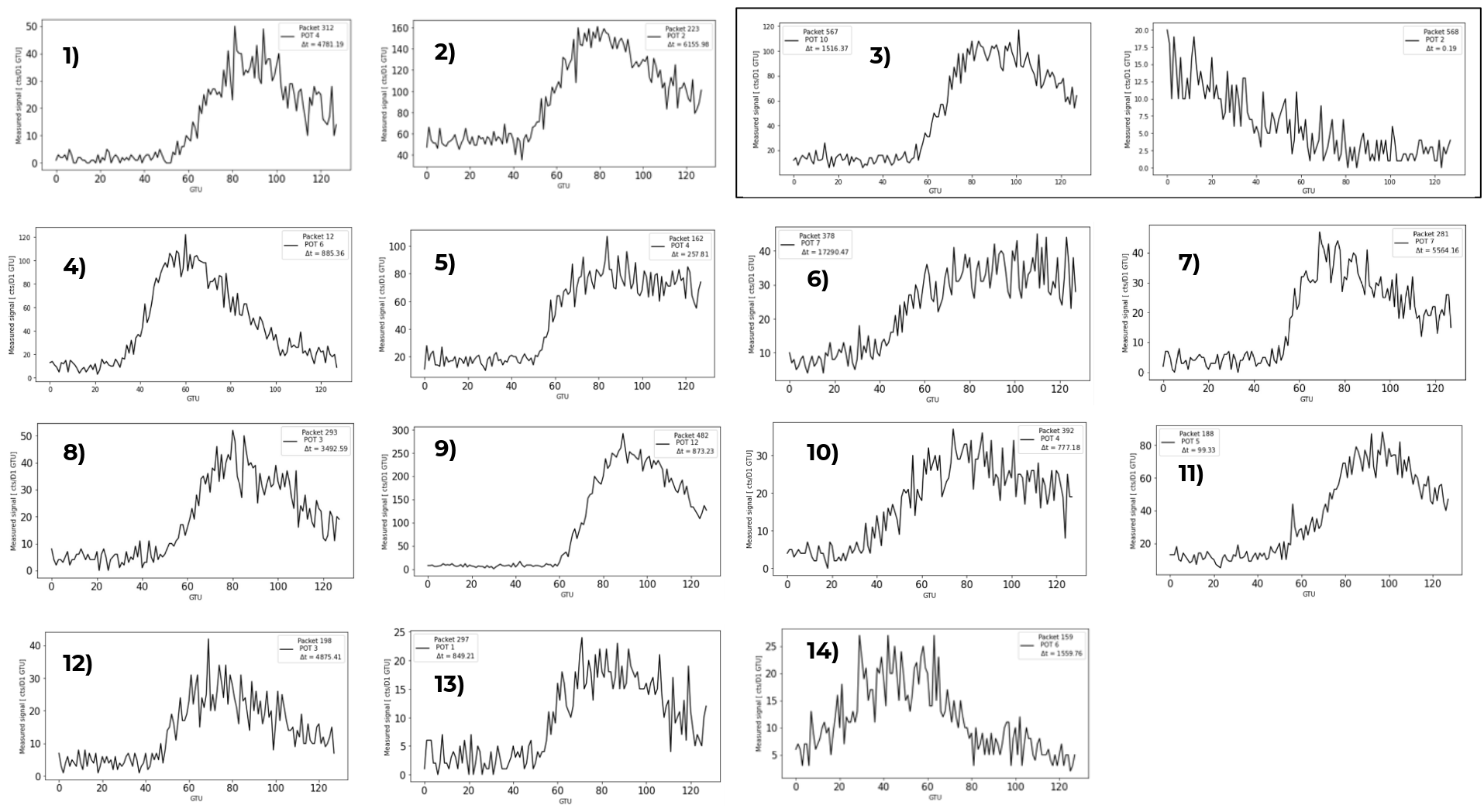}
\caption{Ligtcurves of the 14 SLTs. Event 3 is the only one triggered in two consecutive packets. For events 1, 2, 6, 8, 12, and 13 an atmospheric event has been detected from the exact same position within a few ms after the EAS-like events. Event 4 is the one already shown on the right side of Fig.~\ref{Fig:DCR_and_EAS_like}. See the text for more details.}
\label{Fig:EAS_like_events__lightcurves}
\end{figure}

As mentioned above, we call Short Light Transients any flashing signal lasting no more than 200~$\mu$s that are not originated from a ground flasher. The visual inspection of the dataset identified 14  SLT candidates, whose location is reported in Fig.~\ref{Fig:Map} in blue. In Fig.~\ref{Fig:EAS_like_events__lightcurves} the lightcurves of the events are shown. Usually, these events present a lightcurve resembling a bi-gaussian shape, with a relatively long signal not fully contained in one packet. None of those 14 events present an apparent movement in the focal plane but appear as a stationary light confined in a small cluster of pixels, switching on and reaching the maximum before starting to fade away. It is possible that more SLTs are present in the data since the analysis has been performed manually. A more in-depth analysis of the dataset will be performed once the data currently on the ISS will be available on ground. 

The origin of those fast flashing lights is still under study, but it seems safe to assume that at least some of them are linked with the thunderstorm activity in the atmosphere. For 6 SLT events, in fact (magenta points in Fig.~\ref{Fig:Map}), Mini-EUSO detected an atmospheric event in the exact same pixels shortly after the SLT.  The time interval between the two events can be less than 1~ms (events number 2 and 8 in Fig.~\ref{Fig:EAS_like_events__lightcurves}), between 1 and 4~ms (events number 1 and 13) or much longer (76~ms for event 6 and more than 200~ms for event 12). We are currently working on the identification of those 6 atmospheric events, that appear to belong to the class of Transient Luminous Events (TLEs) rather than to be more common thunder strikes. We are also investigating any possible correlation between these 6 or other Mini-EUSO atmospheric events with Terrestrial Gamma-ray Flashes (TGFs), which are known to be linked to thunderstorm activities~\cite{ASIM_TGFs}. 

The cosmic origin can be excluded with certainty for all the events, also the ones without a follow-up TLE, by a comparison of the focal plane footprint and the lightcurve duration with the simulations shown in Fig.~\ref{fig:ESAF_simulation_Mini_EUSO}. All the 14 events classified as SLTs appear in the focal plane as a small cluster of bright pixels, not too different from the footprint of the 50$^\circ$ zenith angle simulation in Fig.~\ref{fig:ESAF_simulation_Mini_EUSO}. Their lightcurve is however much longer, with a minimum duration of $\sim$80~GTUs ($\sim$200~$\mu$s) for event number 4 and 14, not too distant from the lightcurve of the 80$^\circ$ zenith angle event. For all the events it is therefore not possible to reproduce at the same time the lightcurve shape and duration and the focal plane footprint with a UHECR simulation. This argument by itself is enough to exclude their UHECR origin.

\section{Conclusions}
The ultimate goal of the JEM-EUSO program~\cite{EUSO-Program} is the detection and reconstruction %(in terms of arrival direction, energy and mass composition) 
of UHECRs from space with improved statistics with respect to the current generation of UHECR detectors. In this framework, Mini-EUSO has proved to be an invaluable instrument to address the detection capability of a future, larger space-based detector, like the future POEMMA~\cite{poemma}. Mini-EUSO has measured the terrestrial UV background with unparalleled precision~\cite{UV_maps} and its data have been used to estimate the exposure of a future space-based UHECR detector~\cite{Minieuso_exposure}. The detection of events resembling the duration and the intensities of EAS-induced signals is one of the last missing pieces of the puzzle to prove that a future large-scale space-based detector can significantly contribute to the advancement of the UHECR field. 

The detection of SLTs from Mini-EUSO, in fact, proves that the JEM-EUSO technology can detect UHECRs from space as they show similarities in terms of light profile, intensity, duration, and pixel pattern on the focal surface of the instrument, even though all these characteristics do not match at the same time for a single event. More importantly, Mini-EUSO showed that those events can not be mistaken for real EAS-induced signals, and therefore do not represent a problem for future observations. That confirms the strength of the JEM-EUSO detection principle. Furthermore, a JEM-EUSO or POEMMA-like instrument has a pixel spatial resolution in area $\sim$100 times more refined, and the point-like origin of a flasher-like event would be much more clearly identified and excluded even in the absence of its repetitive pattern.

In addition to that, the ability of Mini-EUSO to detect and study atmospheric phenomena like the ones linked to SLTs is unique and beyond the capabilities of any other atmospheric detector. That makes Mini-EUSO not only a fundamental step in view of the next generation of UHECR detectors but also a unique instrument for the atmospheric science field.

\section*{Acknowledgements}
%The authors acknowledge all members of the JEM-EUSO Collaboration, especially the Mini-EUSO team.
This work was supported by the Italian Space Agency through the agreement n. 2020-26-Hh.0, by the French
space agency CNES, and by the National Science Centre in Poland grants
2017/27/B/ST9/02162 and 2020/37/B/ST9/01821. This research has been
supported by the Interdisciplinary Scientific and Educational School of Moscow
University ``Fundamental and Applied Space Research'' and by Russian State
Space Corporation Roscosmos. The article has been prepared based on research
materials collected in the space experiment ``UV atmosphere''. We thank the
Altea-Lidal collaboration for providing the orbital data of the ISS.

\bibliography{my-bib-database}

\newpage
{\Large\bf Full Authors list: The JEM-EUSO Collaboration\\}
%{\scriptsize (author-list as of July 15th, 2023 with reorganized affiliations)} \hspace{0.6cm}
%{\scriptsize (version  \today{} \currenttime{})}
%\vspace*{0.5cm}

\begin{sloppypar}
{\small \noindent
S.~Abe$^{ff}$, 
J.H.~Adams Jr.$^{ld}$, 
D.~Allard$^{cb}$,
P.~Alldredge$^{ld}$,
R.~Aloisio$^{ep}$,
L.~Anchordoqui$^{le}$,
A.~Anzalone$^{ed,eh}$, 
E.~Arnone$^{ek,el}$,
M.~Bagheri$^{lh}$,
B.~Baret$^{cb}$
% ...
}
\end{sloppypar}
\vspace*{.3cm}

%%\newpage
{ \footnotesize
\noindent
% Algeria - 2 institutes
% ...
$^{cb}$ Universit\'e de Paris, CNRS, AstroParticule et Cosmologie, F-75013 Paris, France\\
% ...
$^{ed}$ Istituto Nazionale di Fisica Nucleare - Sezione di Catania, Italy\\
% ...
$^{eh}$ INAF - Istituto di Astrofisica Spaziale e Fisica Cosmica di Palermo, Italy\\
% ...
$^{ek}$ Istituto Nazionale di Fisica Nucleare - Sezione di Torino, Italy\\
$^{el}$ Dipartimento di Fisica, Universit\`a di Torino, Italy\\
% ...
$^{ep}$ Gran Sasso Science Institute, L'Aquila, Italy\\
% Japan - 7 institutes 
% ...
$^{ff}$ Nihon University Chiyoda, Tokyo, Japan\\ 
% ...
$^{ld}$ University of Alabama in Huntsville, Huntsville, AL, USA\\
$^{le}$ Lehman College, City University of New York (CUNY), NY, USA\\
% ...
$^{lh}$ Georgia Institute of Technology, USA\\
% ...
}

%\begin{thebibliography}{99}
%\bibitem{...}
%....

%\end{thebibliography}

%% Full authors list (ONLY FOR COLLABORATIONS)
%\clearpage
%\section*{Full Authors List: \Coll\ Collaboration}
%
%\noindent \textbf{Note comment afterwards:} Collaborations have the possibility to provide an authors list in xml format which will be used while generating the DOI entries making the full authors list searchable in databases like Inspire HEP. \\
%
%\scriptsize
%\noindent
%first.author$^1$, 
%second.author$^2$, 
%third.author$^3$ % .... more names
%and 
%last.author$^{n}$ \\
%
%\noindent
%$^1$first.affiliation.
%$^2$second.affiliation. % .... more affiliation
%$^{m}$last.affiliation.

\end{document}